# Transient deformations in the Suez-Sinai area from GPS observations

Federica Riguzzi,[1] Antonio Piersanti,[1] Grazia Pietrantonio,[1] and Salah M. Mahmoud[2]

[1]Istituto Nazionale di Geofisica e Vulcanologia, Rome, Italy

[2]National Research Institute of Astronomy and Geophysics, Helwan, Egypt

**SUMMARY**

We analyze data from four GPS campaigns carried out between 1997 and 2002 on a network of 11 sites in the Suez-Sinai, the area of collision between the African and the Arabian plates. This is the key area to understand how and in which way Sinai behaves like a sub-plate of the African plate and the role played between seismic and geodetic (long term) deformation release.

Our analysis shows that, on average, the Suez-Sinai area motion (in terms of ITRF00 velocities) matches African plate motion (NNR-NUVEL-1A model).

However, the baseline length variations show transient deformations in Sinai and across the Gulf of Suez, reaching up a maximum value of about 1.5 cm in five years.

Since current geodynamical models do not predict significant tectonic deformation in this area, we worked under the hypothesis that a contribute may be due to post-seismic relaxation. Under this hypothesis, we compared the baselines length variations with the post-seismic relaxation field associated with five major local earthquakes occurred in the area, testing two different viscoelastic models. Our results show that the transient deformations are better modelled for viscosity values of $10^{18}$ Pa s in the lower crust and $10^{20}$ Pa s in the asthenosphere.

However, since the modelled post-seismic effect results modest and a certain amount of the detected deformation is not accounted for, we think that an improved modelling should take into account the lateral heterogeneities of crust and upper mantle structures.

**Key-words:** Gulf of Suez**,** Sinai, GPS, deformation, post-seismic relaxation

1. INTRODUCTION

The interaction between the African and the Arabian plates displays a unambiguous pattern south of the Sinai peninsula where the opening of the Red Sea rift takes place. Sinai peninsula is a key area for understanding the role played among the Suez rift, the Aqaba fault and the Red Sea rift. Although many authors are discussing the opening of the Suez Gulf, we think that the kinematics engine of the Suez area seems much more complicated than a simple (though low) tensional tectonics. This consideration comes out from the fact that the commonly accepted geodynamical view of the area hardly reconciles with GPS observations and recent focal mechanism solutions. Piersanti et al. [2001] recently enhanced the potential role played by the post-seismic transient deformation with respect to the long-term tectonic deformation.

From November 1997 to May 2002 four GPS surveys were carried out in the Suez-Sinai area on a network of 11 sites (Figure 1), to shed light on its tectonic processes. We reanalyze GPS data from the 1997 and 1998 campaigns [Riguzzi et al., 1999] together with new GPS data from the 2000 and 2002 campaigns. We present ITRF00 horizontal velocities estimated by the GPS time series of each site, after processing the network at regional scale with some selected IGS sites. Thereafter, our attention is focused on the time series of the baseline length variations, since they exhibit significant transient trends different from site to site. In the end, we want to test which could be the contribute due to the post-seismic viscoelastic relaxation to the detected deformations.

2. TECTONIC SETTING

The tectonics of the Suez-Sinai area is dominated by the active boundary between the African and the Arabian plates, which are separating one from the other along the Red Sea rift.

Geological and seismological studies evidence that the Sinai peninsula behaves like a sub-plate of the African plate which accommodates the main motion of the Aqaba left-lateral transform fault (about 8-9 mm/y) [Le Pichon & Gaulier, 1988] with the supposed low extensional motion of the Suez Gulf (< 1 mm/y) [Steckler et al., 1988; Jackson et al., 1988] in the framework of the Red Sea opening [Ben-Menahem et al., 1976]. The triple junction among the Gulf of Suez rift, the Aqaba fault and the Red Sea rift is located south of Sinai. Current geodynamical models support the idea that the triggered motion along the Aqaba fault has almost entirely replaced the Gulf of Suez opening.

Recent GPS results show spreading rates along the Red Sea varying from $14\pm1$ mm $yr^{-1}$ at 15° N to $5.6\pm1$ mm $yr^{-1}$ at 27° N, just below the triple junction, and left lateral strike slip motion of $5.6\pm1$ mm $yr^{-1}$ along the southernmost segment of the Aqaba fault [McClusky et al., 2003]. A very recent analysis based on continuous GPS observation of Israeli network shows the left-lateral motion of the Dead Sea fault (a northern segment of the Aqaba fault) at a rate of about 2.8 mm $yr^{-1}$ and a slight spreading of the Suez Gulf [Wdowinski et al., 2004].

Seismic activity mainly occurs along the borders of the Sinai sub-plate, in the southern part of the Gulf of Suez and along the Aqaba fault, where it reaches the highest level [Mahmoud, 2003; Salamon et al., 2003]. The largest event recorded in the area is the November 22, 1995 (Mw= 7.2), also known as Aqaba or Nuweiba earthquake. This event reached a VIII maximum intensity with substantial damage over an area of about 1000 $km^2$ and was followed by more than 200 aftershocks with $M_l \geq 4.0$ [Malkawi et al., 1999].

From 1997 to 2002 no relevant seismic activity was recorded by the Egyptian National Seismic Network in the area under investigation, since local magnitudes never exceeded 4.6.

3. DATA PROCESSING

The Sinai data of the four GPS campaigns of 1997, 1998, 2000, 2002 were analyzed using the BERNESE GPS sw v.4.2 [Beutler et al., 2001] together with data from some selected IGS permanent stations (Table 1). We adopted the same processing procedure for each campaign, using IGS precise ephemeredes and following the recommended strategy for regional campaigns. The coordinates and covariance matrices pertaining to each session were adjusted in a final campaign solution by the software NETGPS [Crespi & Riguzzi, 1998]. The adjustment was performed constraining within their rms the ITRF00 coordinates of the IGS sites at the mean epoch of each campaign. Table 2 shows the estimated and threshold $\chi^2$ values testing the correctness of the normal variate observation; the test is passed by an iterative procedure when the estimated $\chi^2$ results less than the threshold value. In Table 2 are also reported the precision of the four adjustments in terms of mean rms computed at 95% confidence level.

The horizontal velocity components (Table 1) were estimated by linear fits to time series (Figure 2). Although currently used, this is not a rigorous way to estimate velocities because it assumes a simplified stochastic model by neglecting spatial correlations among sites; while the fact that we have few observations allow us to exclude temporal correlations, thanks to the time lag among observations [Barzaghi et al., 2004].

In terms of global kinematics, our data show that, on average, the Suez-Sinai area motion (ITRF00 velocities) matches the African plate motion defined by the NNR-NUVEL-1A model [De Mets et al., 1994], even if space geodetic rates result on average slower than those predicted by about 7% (Figure 3).

The motion is generally linear in time, except for the sites located on the Sinai shore of the Suez Gulf.

Subsequently, we focus our attention on local Suez-Sinai kinematics, considering GPS baselines obtained after each campaign adjustment together with their errors at 95% confidence level. We selected baselines stemming from GEMS (Table 3), the central site of the African shore having more GPS survey repetitions and the best fit to ITRF00 linear motion (Table 1, Figure 2). Baseline lengths range about from 28 to 134 km with rms never exceeding 3 mm.

Figure 4 shows the time trend of baseline length variations (Table 3) with respect to the first epoch values. All the variations range within ±0.5 cm along the Suez African shore and ± 1.5 cm across the Gulf. It is noteworthy that the detected variations exhibit a larger magnitude (about $10^{-7}$) with respect to the annual scale effect (about $10^{-8}$) recognized in time series of baseline lengths between IGS stations [Pietrantonio, 2002]. Consequently, we want to investigate which could be the contribute of post-seismic relaxation to the significant baseline length variations.

### 4. POST-SEISMIC RELAXATION MODELS

Previous works, based on smaller and less refined dataset [Piersanti et al., 2001], evidenced the possible role played by post-seismic viscoelastic relaxation of the lower crust and asthenosphere in assessing the detected deformation field. At this aim, Piersanti et al. [2001] computed the post-seismic deformation field associated with the 1995 Aqaba earthquake, the major event in that area in the last decades. Here, we repeated the analysis presented in the cited work but since our dataset contains now much more information, we introduced in our simulations also the effects of 5 other significant events occurred in that area in 1993 ($M_W$ = 6.1 and 5.7) 1995 ($M_W$ = 5.3) 1996 ($M_W$ = 5.3) and 2000 ($M_L$ = 4.6), see Table 4. We used the CMT solutions for the source focal parameters [Dziewonski et al., 1997], while for the epicentral locations we prefer to use more precise data coming from a local seismic network [Salomon et al., 2003].
We remind that our model of post-seismic deformation takes into account the effects of spherical geometry self gravitation, linear viscoelastic rheology of the asthenosphere and lower crust, physical and chemical discontinuity between the elastic shallow layer (crust) and the two viscoelastic layers (lower crust and asthenosphere) [Piersanti et al., 1995; 1997]. The adoption of a spherical model is likely to be not meaningless, since the area under examination has a linear extension of some hundreds of km (see Nostro et al., [1999]). The solution is retrieved semi-analytically, through a spherical harmonic decomposition of the physical observables. The reader is referred to the previously cited works for a detailed description of the model implementation and solution method. Piersanti et al. [2001] tested 6 different layerings. Since two of them (model 2 and 3 from table 1 of Piersanti et al., 2001) furnished sensibly better results than the others, in the present simulations we used these two particular profiles. They are composed by an elastic crust (thickness 15 and 20 km respectively), a viscoelastic lower crust (thickness 15 km, viscosity $10^{18}$ Pa s) and a viscoelastic asthenosphere (thickness 170 and 165 km, viscosity $10^{18}$ and $10^{20}$ Pa s, respectively).

Figure 4 shows the time series of GPS baseline length variations vs. the post-seismic relaxation models 2 and 3. The variations predicted in 5 years range between -2 cm and +1 cm, except for the baseline GEMS-CATH, where the predicted contraction reaches the highest level. It has to be underlined that the model curves of Figure 4 are not retrieved by inversion but by forward-type modelling, so they do not represent any fit of data, but refer to a completely independent data set (from seismology).

With the aim of evaluating which model agrees better with the GPS baseline variations, we performed a Fisher test (Koch,1988). The F experimental value is $F_e = 3.24$ while the theoretical is $F_{28,28} = 2.46$ at 99% significance level, therefore the models are not equivalent. From statistics we should prefer model 3, since the mean deviations are the smallest.

In our opinion, though it is noteworthy that GPS and post-seismic derived deformations are comparable in magnitude (mean deviations are within 1 cm) and time trend, on the other hand, the comparison among relaxation models adequacy to the data and that of the no-change model shows that the contribute due to relaxation is small (see Table 5).

## 5. DISCUSSION AND CONCLUSIONS

The utility of GPS to detect crustal deformations is widely recognized, so that permanent GPS arrays are currently used for this purpose all over the world. Anyway, non-permanent networks play important roles in control surveys, where establishing permanent GPS arrays is too expensive besides logistically and politically hard.

Taking into account what stated above, we have analyzed GPS data coming from four surveys on a non-permanent network of 11 sites in the Suez-Sinai area, carried out from 1997 to 2002. The main conclusions of our analysis, focused on ITRF00 site velocities and baseline length variations across the Suez channel, are

1) the estimated ITRF00 GPS velocities are consistent with those predicted by the NNR-NUVEL-1A, though smaller by about 7%. This implies that, on average, the Sinai-Suez area matches the African plate motion defined by NNR-NUVEL-1A and the extension in the Suez Gulf results at present not active. Consequently, GPS data do not allow us to distinguish Sinai as sub-plate of the African plate and most of the motion would concern Arabia with respect to Africa, in agreement with Sella et al. [2002] and McClusky et al. [2003]. The hypothesized deceleration of the African plate enhanced by Sella et al. [2002], McClusky et al. [2003] and Calais et al. [2003] is here supported by the detected difference between GPS and NNR-NUVEL-1A velocities, though near the significance level;

2) the significant transient trend of the baseline length variations across the Suez Gulf exhibits in our opinion a possible signature of seismic origin in terms of post-seismic viscoelastic relaxation, in agreement with the role enhanced by Piersanti et al. [2001]; under this hypothesis the assessed transient deformations are better modelled for viscosity values of $10^{18}$ Pa s in the lower crust and $10^{20}$ Pa s in the asthenosphere.

Anyway, we showed that the post-seismic contribute given by our models is not so large if compared with the no-change model (Table 5). The detected discrepancies between the post-seismic predicted and the GPS observed deformations could be due to the model assumption of lateral homogeneity of the medium, in spite of the complex tectonics of the area. Even if the knowledge of the crustal and upper mantle structure of the region is very limited, the Gulf of Suez rift would display sharp lateral heterogeneities between continental and oceanic crust (Steckler, 1985; Steckler et al., 1998). Under this light should be considered the largest deviations exhibited by CATH, the highest site (1350 m above s. l.) located in the thicker continental crust of Sinai.

We believe that in the next future new perspectives on the geodynamics of the Suez-Sinai area will be opened by combining all the existing GPS solutions across and around this area and improving local crustal models with lateral heterogeneities.

## ACKNOWLEDGMENTS

We thank E. Boschi, A. Amato and A. Tealeb for supporting this work. We are grateful to the personnel of NRIAG involved in the GPS campaigns.

The maps were created by using the public domain Online Map Creation (http://www.aquarius.geomar.de/omc/make_map.html) based on the Generic Mapping Tools (GMT) software [Wessel and Smith, 1995].

**Table 1**

GPS sites and observation years, ITRF00 horizontal velocities and NNR-Nuvel-1A predicted velocities

| Site | Campaign year | | | | Geographic Coordinate | | ITRF00 velocity | | Nuvel1A velocity | |
|------|----|----|----|----|------|------|------|------|------|------|
|      | 97 | 98 | 00 | 02 | Lat. (°) | Long. (°) | $V_N$ $\left(\frac{mm}{yr}\right)$ | $V_E$ $\left(\frac{mm}{yr}\right)$ | $V_N$ $\left(\frac{mm}{yr}\right)$ | $V_E$ $\left(\frac{mm}{yr}\right)$ |
| CATH | ▒ | ▒ | ▒ | ▒ | 28.64 | 34.00 | 15.1±0.5 | 24.8±3.0 | 19.5 | 25.0 |
| DAHA | ▒ | ▒ | ▒ | ▒ | 28.53 | 34.47 | 19.9±1.5 | 21.6±1.7 | 19.5 | 25.1 |
| DERB | ▒ | ▒ | ▒ | ▒ | 28.63 | 33.40 | 17.9±2.9 | 22.5±1.3 | 19.6 | 24.9 |
| ELAT | ■ |   |   |   | 29.51 | 34.92 | ITRF00 | | | |
| GARB | ▒ | ▒ | ▒ | ▒ | 28.16 | 33.23 | 19.0±2.0 | 23.6±0.8 | 19.6 | 24.9 |
| GEMS | ▒ | ▒ | ▒ | ▒ | 27.69 | 33.49 | 19.0±1.1 | 23.5±0.7 | 19.6 | 25.0 |
| HURG | ▒ | ▒ | ▒ | ▒ | 27.25 | 33.83 | 18.1±1.7 | 22.7±0.5 | 19.6 | 25.1 |
| KENS | ▒ | ▒ | ▒ |   | 27.96 | 33.88 | 17.0±1.2 | 21.8±3.0 | 19.6 | 25.0 |
| LAMP |   |   |   | ■ | 35.50 | 12.61 | ITRF00 | | | |
| MATE | ■ | ■ | ■ | ■ | 40.65 | 16.70 | ITRF00 | | | |
| NABQ | ▒ | ▒ | ▒ |   | 28.18 | 34.31 | 19.5±1.6 | 23.1±1.8 | 19.5 | 25.1 |
| NICO | ■ | ■ | ■ | ■ | 35.14 | 33.40 | ITRF00 | | | |
| SHAM | ▒ | ▒ | ▒ | ▒ | 27.85 | 34.18 | 18.5±2.0 | 23.8±1.6 | 19.5 | 25.1 |
| SOFI | ■ | ■ |   | ■ | 42.56 | 23.40 | ITRF00 | | | |
| TOUR | ▒ | ▒ | ▒ | ▒ | 28.27 | 33.60 | 17.7±2.7 | 23.9±1.6 | 19.6 | 24.9 |
| ZECK | ■ | ■ | ■ | ■ | 43.79 | 41.57 | ITRF00 | | | |
| ZEIT | ▒ | ▒ | ▒ | ▒ | 27.92 | 33.39 | 19.3±1.9 | 23.1±0.4 | 19.6 | 24.9 |

**Table 2**

Network adjustments: $\chi^2$ global test and mean precision at 95% confidence level

| Campaign | $\chi_t^2$ | $\chi_e^2$ | Mean rms (cm) | |
|----------|-----------|-----------|-----|-----|
|          |           |           | φ | λ |
| 1997.91 | 260.99 | 256.78 | 0.2 | 0.2 |
| 1998.41 | 129.92 | 125.32 | 0.2 | 0.2 |
| 2000.72 | 96.22  | 83.78  | 0.3 | 0.3 |
| 2002.37 | 163.12 | 134.34 | 0.3 | 0.3 |

**Table 3**
Baseline length at initial epoch 1997.91 and variations with rms

| Baseline | 1997.91 (m) | 1998.41 (m) | 2000.72 (m) | 2002.37 (m) |
|---|---|---|---|---|
| Gems-Cath | 116454.601±0.001 | -0.009±0.002 | -0.007±0.002 | - |
| Gems-Daha | 133825.289±0.002 | -0.006±0.003 | -0.004±0.003 | -0.004±0.003 |
| Gems-Derb | 105029.420±0.002 | -0.009±0.003 | -0.017±0.003 | -0.003±0.002 |
| Gems-Garb | 58962.818±0.002 | 0.004±0.004 | -0.003±0.003 | 0.003±0.003 |
| Gems-Hurg | 59267.209±0.002 | 0.003±0.004 | 0.005±0.003 | 0.000±0.003 |
| Gems-Kens | 48911.271±0.002 | -0.006±0.003 | -0.004±0.003 | - |
| Gems-Nabq | 97371.468±0.002 | -0.010±0.004 | -0.004±0.003 | -0.004±0.003 |
| Gems-Sham | 70245.158±0.002 | -0.004±0.003 | 0.009±0.004 | -0.005±0.003 |
| Gems-Tour | 65402.937±0.001 | -0.011±0.002 | -0.015±0.002 | -0.007±0.001 |
| Gems-Zeit | 27678.038±0.002 | 0.000±0.003 | -0.005±0.003 | 0.004±0.003 |

**Table 4**
Seismic events selected for modelling

| Date, time(GMT) | Location Lat. (°) | Long. (°) | $M_L$ | $M_w$ |
|---|---|---|---|---|
| 1993 03 08 a, 12: 43 | 28.78 | 34.57 | | 6.1 |
| 1993 03 08 b, 16:33 | 28.79 | 34.59 | | 5.7 |
| 1995 11 22 a, 04:15 | 28.81 | 34.80 | | 7.2 |
| 1995 11 22 b, 22:16 | 28.55 | 34.72 | | 5.3 |
| 1996 02 21, 04:59 | 28.86 | 34.71 | | 5.3 |
| 2000 06 25, 19:18 | 28.21 | 33.48 | 4.6 | |

**Table 5**
Adequacy of post-seismic and no-change models

| Baseline | Mean deviations (mm) | | |
|---|---|---|---|
| | Model 2 | Model 3 | No-change |
| Gems-Cath | 23 | 8 | 8 |
| Gems-Daha | 10 | 3 | 5 |
| Gems-Derb | 10 | 11 | 10 |
| Gems-Garb | 3 | 3 | 3 |
| Gems-Hurg | 5 | 3 | 3 |
| Gems-Kens | 4 | 3 | 5 |
| Gems-Nabq | 7 | 5 | 6 |
| Gems-Sham | 7 | 7 | 6 |
| Gems-Tour | 7 | 7 | 11 |
| Gems-Zeit | 3 | 3 | 3 |
| **mean** | **8** | **5** | **6** |

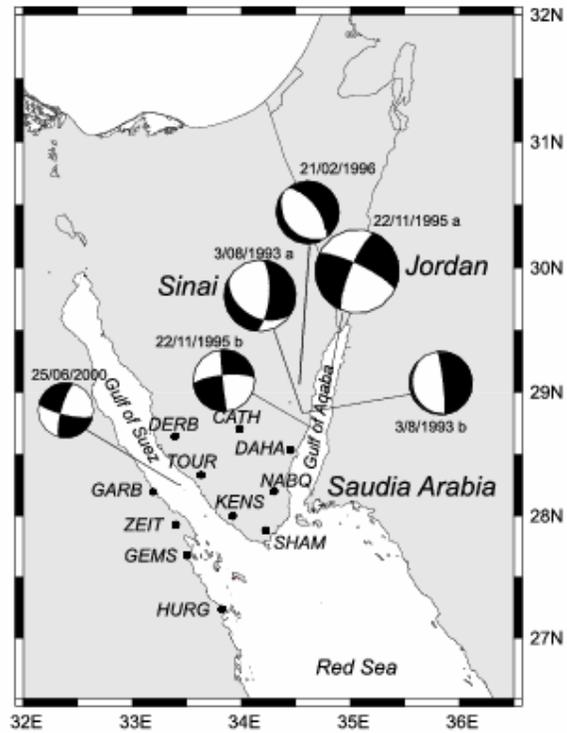

**Figure 1.** The Suez-Sinai GPS network and focal mechanisms selected from the CMT catalogue (Table 4).

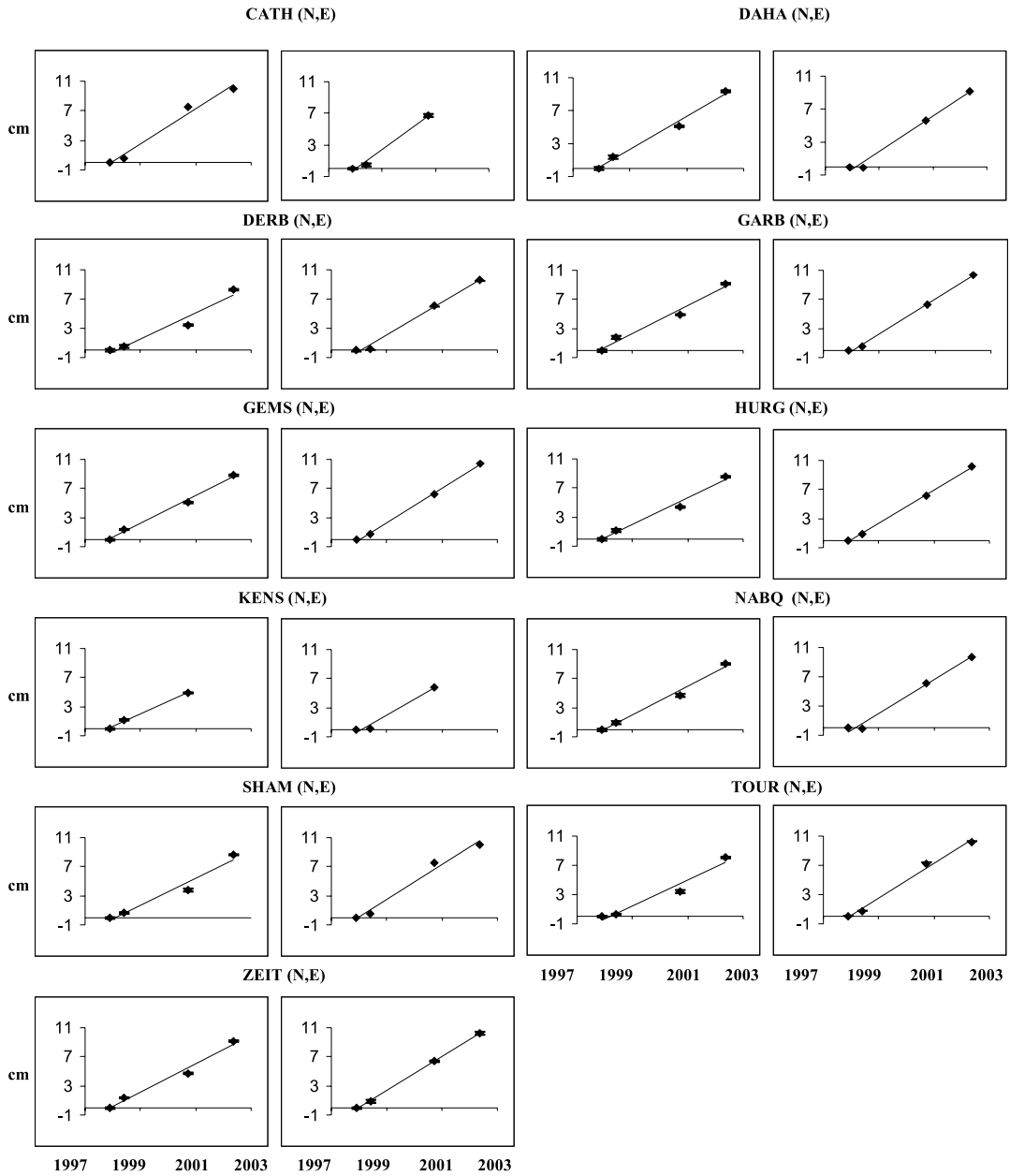

**Figure 2.** ITRF00 coordinate (N, E) time series of the GPS sites (error bars at 95% confidence interval). The straight lines are the weighted linear regression of campaign solutions.

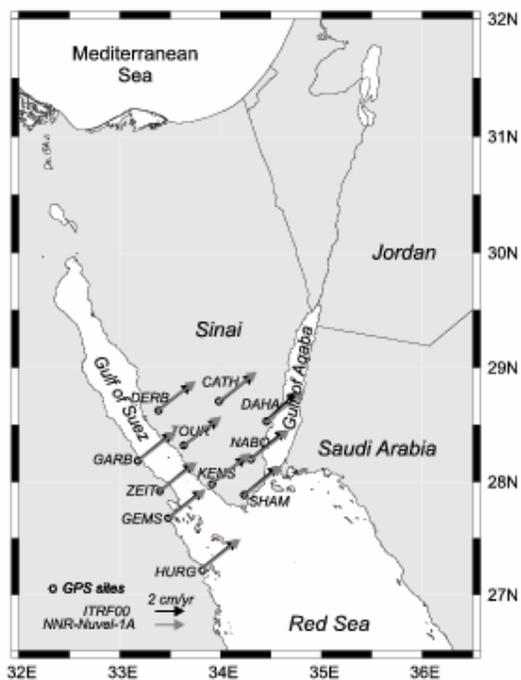

**Figure 3.** NNR-NUVEL-1A (grey) and ITRF00 (black) horizontal velocities.

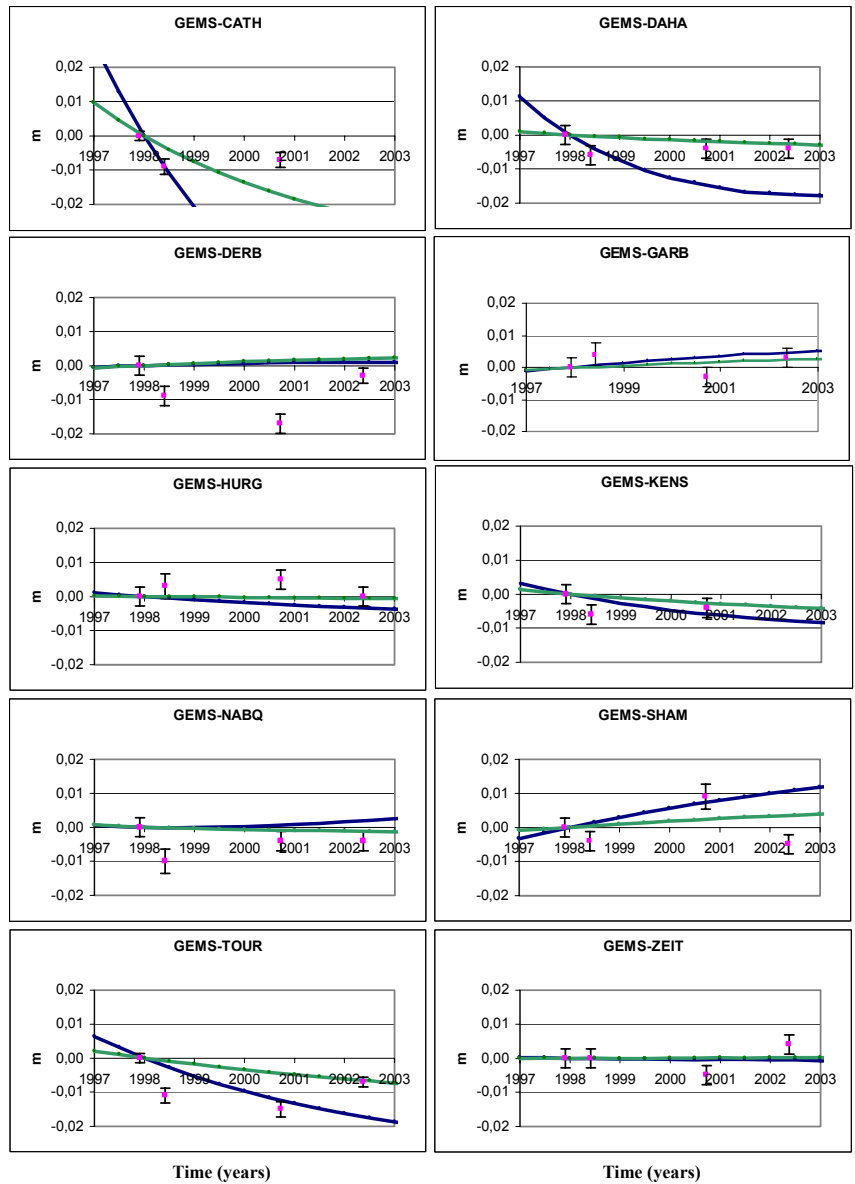

**Figure 4.** Time series of baseline length variations (error bars at 95% confidence interval) vs. post-seismic relaxation models (2 in blue, 3 in green).